\documentclass[mathleft
]{an}
\usepackage{graphicx}
\usepackage{times}
\newcommand {\xmm}{{XMM-\it Newton}\ \ignorespaces}
\begin{document}

\Pagespan{789}{}
\Yearpublication{2006}%
\Yearsubmission{2005}%
\Month{11}%
\Volume{999}%
\Issue{88}%

\title{The origin of the strong soft excess and puzzling iron line complex in Mkn 841}

\author{P.O.\,Petrucci\inst{1} \and G. Ponti\inst{2} \and
  G.\,Matt\inst{3} \and 
L.\,Maraschi\inst{4} \and J.\,Malzac\inst{5} \and M.\,Mouchet\inst{6} \and
C.\,Boisson\inst{6} \and A.\,Longinotti\inst{7} \and K.\,Nandra\inst{8} 
\and P.\,Ferrando\inst{9} \and G.\,Henri\inst{1}}

\institute{Laboratoire d'Astrophysique de Grenoble, BP 43, 38041 Grenoble
  Cedex 9, France \and NAF-IASF Sezione di Bologna, Via Gobetti 101,
  I-40129 Bologna, Italy \and Dipartimento di Fisica, 
  Universit\`a degli Studi ``Roma tre'', via della Vasca Navale 84,
  I-00046 Roma, Italy \and Osservatorio Astronomico di Brera, Via Brera 28,
  02121 Milano, Italy \and Centre d'\'etude Spatiale des Rayonnements
  (CNRS/UPS/OMP), 31028 Toulouse, France  \and LUTH, Observatoire de
  Paris, Section de 
  Meudon, 92195 Meudon Cedex, France \and XMM-Newton Science
  Operations Center, European Space Astronomy Center, ESA, 28080
  Madrid, Spain  \and Astrophysics 
  Group, Imperial College London, Blackett Laboratory, Prince Consort
  Road, London SW7 2AW \and Service d'Astrophysique, DSM/DAPNIA/SAp,
  CE Saclay, 91191 Gif-sur-Yvette Cedex, France   }

\titlerunning{Soft excess and Iron line complex in Mkn 841}
\authorrunning{P.O. Petrucci et al.}

\received{30 May 2005}
\accepted{11 Nov 2005}
\publonline{later}


\abstract{Mkn 841 has been observed during 3 different periods (January
2001, January 2005 and July 2005) by XMM-Newton for a total cumulated
exposure time of $\sim$108 ks. We present in this paper a broad band
spectral analysis of the complete EPIC-pn data
sets. These observations confirm the presence of the strong soft excess
and complex iron line profile known to be present in this source since a
long time. They also reveal their extreme and puzzling spectral and
temporal behaviors. Indeed, the 0.5-2 keV soft X-ray flux decreases by
a factor 
  3 between 2001 and 2005 and the line shape appears to be a mixed of
  broad and narrow components, both variable but on different
  timescales. The broad-band 0.5-10 keV spectra are well described by a
model including a primary power law continuum, a blurred photoionized
reflection and a narrow iron line, the blurred reflection fitting
self-consistently the soft excess and the broad line component. The
origin and nature of the narrow component is unclear.}

\maketitle

\begin{table*}
\caption{Observation epochs, exposure and mean count rates.\label{log}}
\begin{center}
\begin{tabular}{lccccc}
\hline
 & OBS 1 & OBS 2 & OBS 3 & OBS4 & OBS5 \\
\hline
Start date & 2001-01-13 & 2001-01-13 & 2001-01-14 & 2005-01-16 & 2005-07-17 \\
 & (05h20m55s UT) & (09h33m50s UT) & (00h52m28s UT) & (12h38m21s UT) & (06h38m03s UT)\\
Exposure (s) & 8449 & 10900 & 13360 & 45982 & 29071\\
Cts.s$^{-1}$ PN & 18.0 & 22.2 & 21.8 & 5.6 & 7.2 \\
\hline
\end{tabular}
\end{center}

\end{table*}

\section{Introduction}
Mkn 841 is a bright
Seyfert 1 galaxy (z=0.0364, Falco et al. 2000), one of the rare
Seyfert 1 detected by OSSE 
at more than 3 $\sigma$ (Johnson 1997). It is known for its large
spectral variability (George et al. 1993; Nandra et
al. 1995), its strong soft excess (this was the first object
where a soft excess was observed, Arnaud et al. 1985) and its
variable iron line (at least on a year time scale, George et
al. 1993). Here we report the data analysis of the complete set of
XMM-Newton observations of this source, focusing on the EPIC-pn
instrument. These observations confirm the presence of the soft excess
and iron line complex and reveal their extreme and puzzling spectral
and temporal   behaviors.\vspace*{-0.2cm}

\section{Data}
\vspace*{-0.2cm}
\begin{figure*}[t]
\includegraphics[width=\textwidth,angle=0]{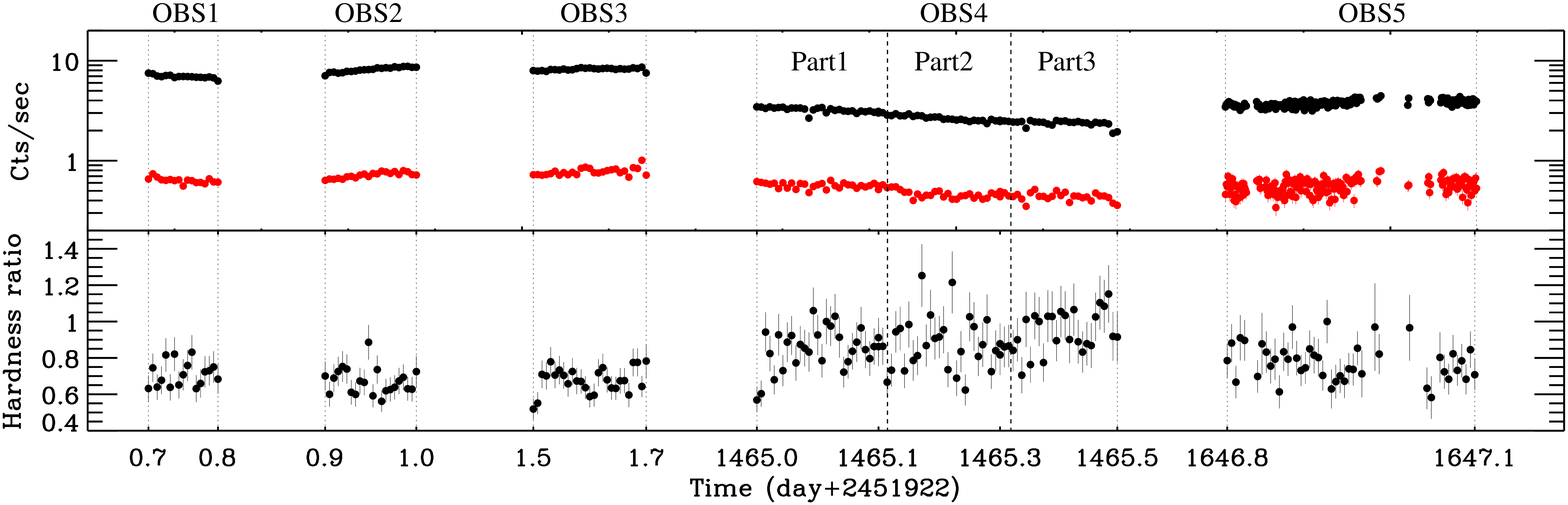}
\caption{Upper panel: Total (0.5-10 keV, upper curve) and hard (3-10
  keV, lower curve) X-ray light curves of the different EPIC-pn
  observations of Mkn 
  841. The broad band flux is clearly dominated by the soft ($<$ 3 keV)
  bands. The soft X-rays count rate varies by a factor 4 in 4 years while
  the hard X-ray one varies by only $\sim$ 60\%. On the other hand flux
  variability up to $\sim$50\% is also observed on tens of ks. The
  hardness ratio is plotted on the lower panel.}
\label{lcfig}
\vspace*{-0.5cm}
\end{figure*}
Mkn 841 has been observed 5 times during 3 different periods (January 2001,
  January 2005 and July 2005) by XMM-Newton for a total cumulated
  exposure time of $\sim$108 ks. Table \ref{log} gives a summary of
  the different XMM-Newton 
pointings with the corresponding dates, exposures and count rates. For
the spectral analysis we have divided OBS4 (jan. 2005) in 3 parts
  (noted part1, part2 
and part3 in the following) of about 15 ks each.

The EPIC-pn camera was always operated in Small
Window mode, with thin aluminium filters to reject visible
light.  The EPIC-pn event files were reprocessed from the ODF data
files using the 
{\it epchain} pipeline tasks of the XMM Science
Analysis System (SAS version 6.5) and using the most
updated version of the public calibration files. These event files were
then filtered for good time intervals following the ``recipe'' given in
the XMM handbook (v2.01 23 July 2004, Sect. 5.2.4).  The source
spectra and light curves were 
built from photons detected within a 40 arcsec extraction window centered
on the source. X-ray events 
corresponding to patterns $\le$ 4 were selected. The background was
estimated within a window  
of the same size from an offset position.

In the following, all errors refer to 90\% confidence level for 1
interesting parameter ($\Delta\chi^2$=2.7).\vspace*{-0.2cm}

\section{Light curves, hardness ratios}
\label{sectlc}
\vspace*{-0.2cm}
We have plotted on top of Fig. \ref{lcfig} the 0.5-10 keV EPIC-pn count
rate light curves of the different \xmm observations of Mkn
841 as well as the hardness ratio (5-10)/(3-5) at the bottom. The time
binning is 500 sec. The 0.5-10 keV count rate decreases by a factor
$\sim$4 in 4 years while the hardness ratio increases, reaching maximum
values during OBS 4. The 3-10 keV count rate has been also plotted 
in Fig. \ref{lcfig}. It shows variations of $\sim$ 60\% implying that the
0.5-10 keV count rate variability is dominated by the soft ($<$ 3 keV)
X-ray variability, at least on long time scale.  Smooth soft {\it and}
hard flux variabilities up to $\sim$50\% are also visible on tens of
ks.\vspace*{-0.2cm}

\section{Simple spectral analysis}
\vspace*{-0.2cm}
We use first very simple spectral components to
reproduce the observed features. We use a power law for the continuum and
a gaussian for the iron line. We fit the data above 3 keV first. Then we fix the different
parameters and include the data below 3 keV down to 0.5 keV and we add a
simple black body component to model the soft excess.  
\begin{figure}[!t]
\begin{center}
\includegraphics[width=\columnwidth]{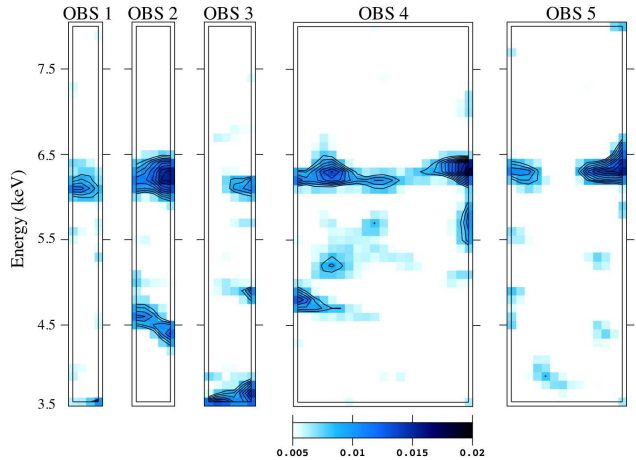}
\caption{Map excess of the different XMM observations \label{mapexcess}}
\end{center}
\vspace*{-0.8cm}
\end{figure}
\begin{figure*}
\begin{tabular}{cc}
\includegraphics[width=0.97\columnwidth]{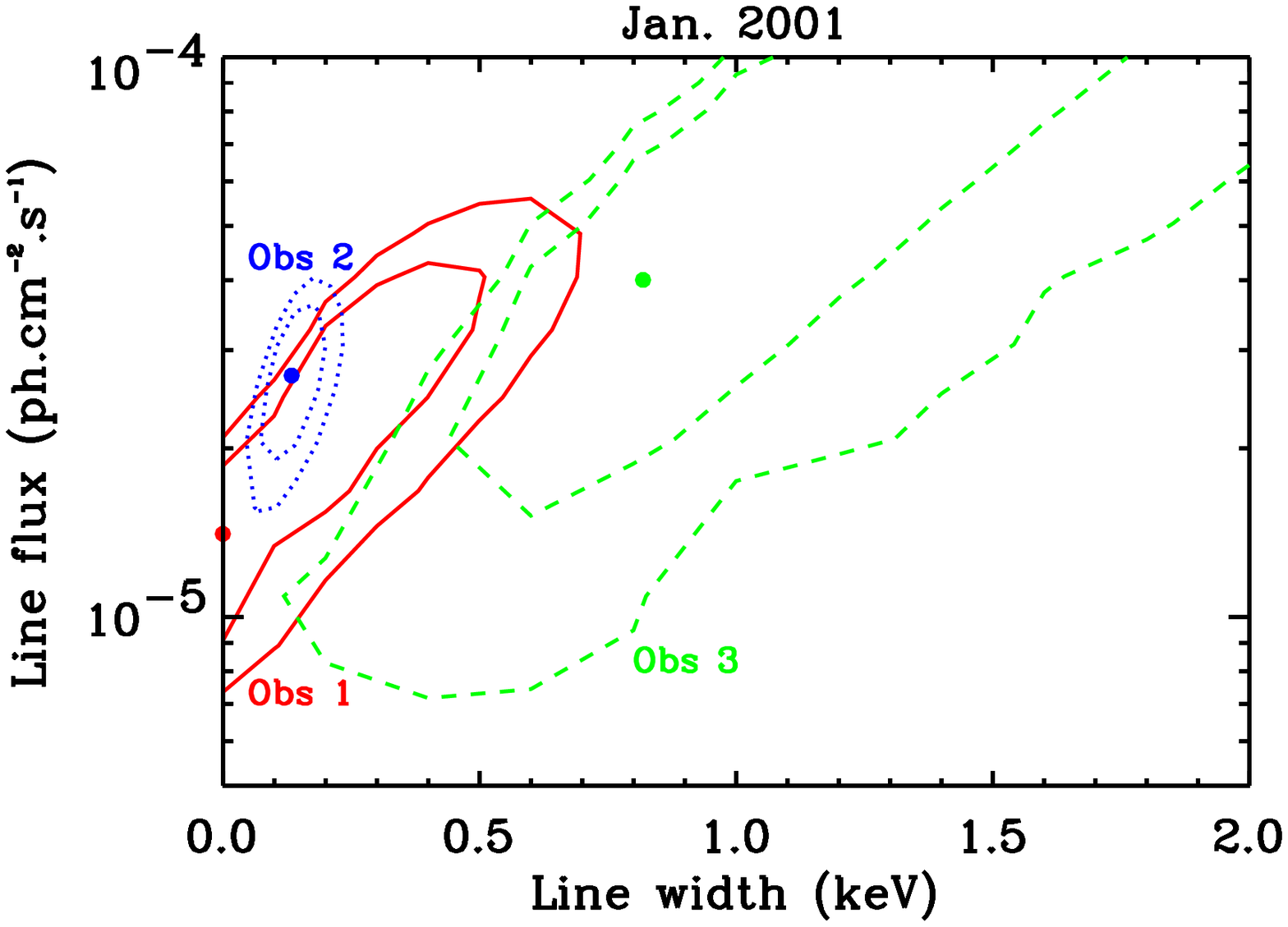}&
\includegraphics[width=0.97\columnwidth]{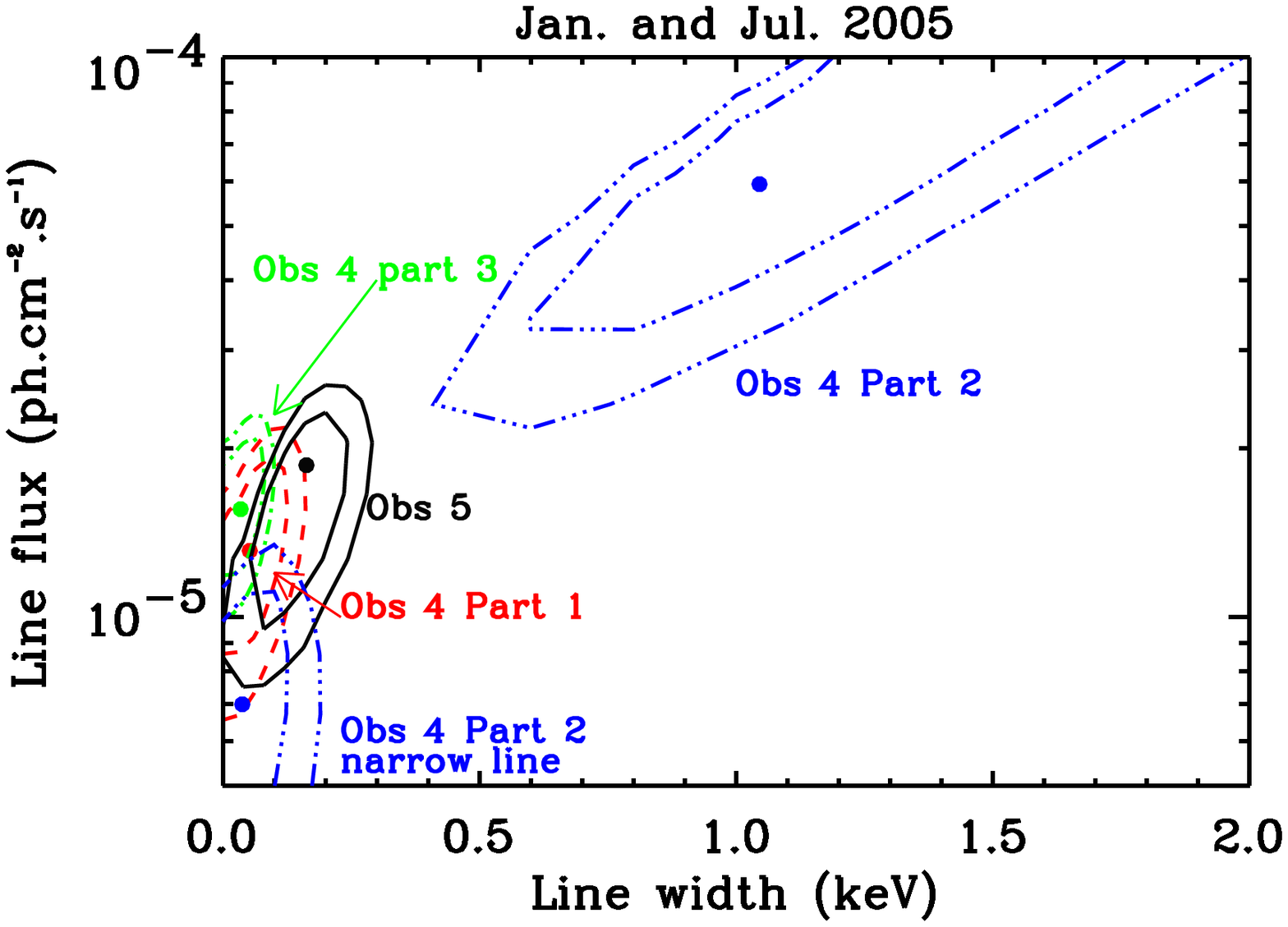}\\
\end{tabular}
\caption{Contour plots (68 and 90 \%) of the line width vs. line flux
  for the 3 observations of Jan 2001
  (left) and the 3 parts of Jan. 2005 (right). In both cases, the model
  includes a simple power law and a gaussian line. We have also over
  plotted on the right the contour plots of OSB 5 
  as well as the contours of the narrow component detected in part 2 of OBS
  4.\label{contplot}}
\vspace*{-0.5cm}
\end{figure*}
The best fits are not satisfactory for all the observations due
to discrepancies present below 3 keV. Anyway different
remarks, weakly affected by the goodness of the fit, can already been
done:\\
1) The spectral variability appears to be due to both variable high
  energy (above 3 keV) continuum and variable soft excess. Indeed the
  photon index varies by $\sim$ 0.5 between 2001 and 2005, while the soft
  X-ray flux, below 3 keV, decreases by a factor 3, i.e. more than the
  flux variability (of $\sim$ 60\%) expected if the power law continuum
  was the only variable component.\\
2) The photon index reaches values as small as 1.3 during OBS 4,
  unusually hard for a Seyfert X-ray spectrum.\\
3) The best fit requires a narrow line in all the observations but OBS
  3 and part 2 of OSB 4 where a broad component is found. The apparent 
  variability of the line width on short time scale in 2001 was
  already noted by Petrucci et 
  al. (2002) and Longinotti et al. (2004) and we observe the same variability
  behavior during the observation of Jan. 2005. This variability is clearly
  visible on the map excess reported  in Fig. \ref{mapexcess}
  (and produced following the method describe in
  Iwasawa et al. 2004). The contour
plots, at 68 and 90 \% confidence level, of the line flux vs. line width
for the different observations are plotted in  Fig. \ref{contplot}.
%
%
%

\begin{figure}[b!]
\includegraphics[height=\columnwidth,angle=-90]{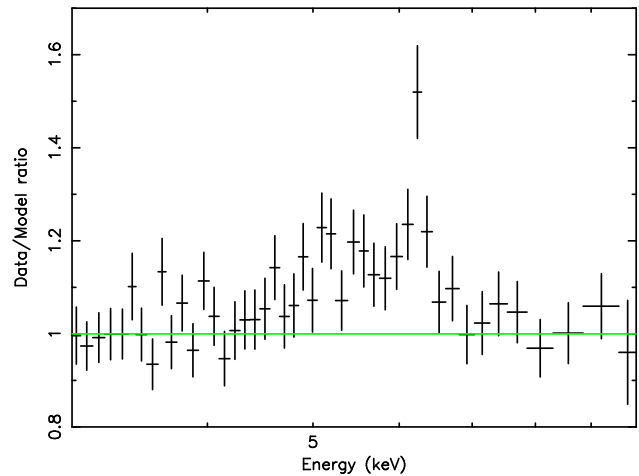}
\caption{Ratio data/model for OBS 4. The model is a simple power law
fitted between 3 and 10 keV, ignoring the 4-7 keV range. Broad and narrow
components are clearly visible.\label{broadline}}
\vspace*{-0.5cm}
\end{figure}

Somes differences exist however between the observations of Jan. 2001 and Jan. 2005. First of all, we note an increase
of the line flux during part 2 of OBS4, a factor 2-3 larger than in
part 1 and 3 
while in 2001 the line flux kept constant between the narrow and broad
component. More importantly we detect some residuals of a narrow
component near 6.4 keV in part 2 of OBS 4. There are also some residuals
near 5-6 keV in part 1 and 3 suggesting the presence, in these cases,
of a broad 
component. To check this, we have added a second gaussian to the model. A
narrow component is indeed detected in part 2 while broad ones are found
in part 1 and 3.   The addition of this second line is significant at
more than 99.7\%, 93\% and 98.9\% (following the F test) for part 1, 2
and 3 respectively. Thus the line complex during OBS 4 appears to be a mix of broad and
narrow components. This is also illustrated in Fig. \ref{broadline} where
we have plotted the ratio of the complete OBS 4 data spectrum and the
best fit power law obtained fitting only the data above 3 keV but
ignoring the 4-7 keV energy range. The presence of both broad and narrow
components is clearly visible. This contrasts with what we observed in
the other observations (OBS 1, 2 or 3 and OBS 5) where the addition of a
second gaussian line does not improve significantly the fits.

It is also worth noting that the narrow line detected 
in part 2 of OBS 4 has a weak flux but in relatively good agreement with
the fluxes of 
the other narrow components detected in OBS 1 and OBS 2 in 2001, in
part 1 and 3   
of OBS 4 and in OBS 5.  The corresponding contour plot
has been over-plotted on Fig. \ref{contplot}. This would suggest the
narrow line to be produced in a remote material  but the lack of
a narrow 
component in OBS 3 disagrees with this picture. Thus the origin of this
component is unclear. Concerning the broad
components detected in part 1 and 3, they are consistent with each other
and with the broad component of part 2.

We have also tried to fit the 3 parts of OBS 4 with a model of line
emission from a relativistic accretion disk ({\sc{diskline}} model of
{\sc{xspec}}) fixing the accretion disc outer radius to 1000 $r_g$ and
keeping the inclination angle constant between the different spectra. The
fit is good, with a best fit inclination angle of 47$_{-8}^{+5}$
degrees. We note however that the fit with two gaussians is slightly
better for part 3 of OBS 4 compared to the {\sc{diskline}} one
($\chi^2/dof$=131/140 compared to 141/141) thanks to a better fit of the
narrow component. 

\begin{figure}[b!]
\vspace*{-0.5cm}
\includegraphics[width=\columnwidth]{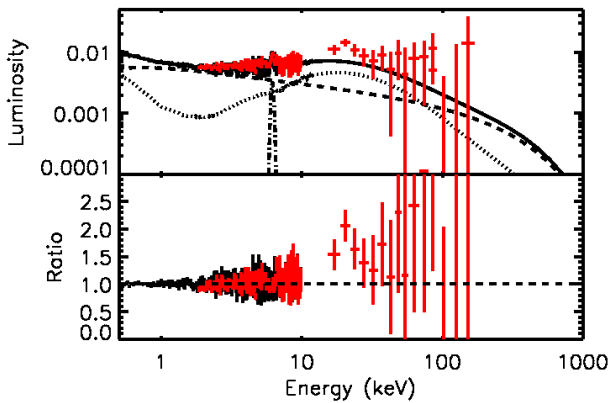}
\caption{Top: The solid line is the best fit blurred ionized reflection
  model of OBS 1, the cut-off power law continuum and blurred ionized
  reflection being plotted in dashed and dotted lines
  respectively. The dot-dashed line is the narrow iron line. The dark (black)
  crosses are the OBS 1 XMM data and the clear 
  (red) crosses are the MECS and PDS data of the simultaneous 100 ks
  BeppoSAX. The PDS/MECS cross normalization was fixed and only the
  XMM-pn/MECS one was let free to vary. Bottom: Data/model ratio
  \label{xmmsax}} 
\end{figure}
\section{A more physical analysis: blurred ionized reflection?}
Recent studies suggest that an appealing explanation for the presence of
strong soft excesses in AGNs could be blurred (photoionized) reflection
from the accretion disc (e.g. Crummy et al. 2005). These authors already
applied this model to the OBS 3 of Mkn 841 with success.  Moreover, in
the cases in which a broad Fe line is clearly detected (such as in
MCG--6-30-15) the model is very robust because the soft excess and broad
Fe line are fitted self--consistently with the same relativistically
blurred reflection model. The characteristics of the line complex in Mkn
841, with the presence of a variable broad line component well fitted by
a blurred ({\sc{diskline}}) profile, and a strong soft excess, suggest
that photoionized reflection may play also an important role in this
source.

To test these assumptions we have used a more physical model with the
following components: 1) a neutral absorption free to vary above the
galactic value 2) a cut-off power law continuum (the high energy cut-off
being fixed to 300 keV), 3) a blurred photoionized reflection and 4) a
narrow line component if needed. For the blurred reflection, we use the
tables of the Ross \& Fabian code (Ross \& Fabian 2005) convolved with a
Laor profile ({\sc kdblur} kernel). This component is expected to
reproduce the soft excess and the broad iron line. The inner radius of
the reflecting accretion disc as well as the disc emissivity law spectra
index were let free to vary in the Laor kernel. The outer radius was
fixed to 400 Schwarschild radii. The iron abundance was also fixed to the
solar one for the computation of the ionized reflection.
\begin{table}[t]
\caption{Best fit results obtained with the blurred ionized reflection
  model. \label{tabkdb}}
\begin{tabular}{lccccc}
\hline
Obs. & $\Gamma$ & $\xi$ & $q$ & $r_{in}$& $\chi^2$/dof\\
 & & & & $r_g$ & \\
\hline
1 & 2.31$_{-0.09}^{+0.07}$ & 190$_{-80}^{+160}$ &8.7$_{-1.4}^{+1.3}$ & 1.3$_{-0.1}^{+0.2}$ & 248/241\\ 
2 & 2.28$_{-0.05}^{+0.10}$ & 117$_{-10}^{+140}$ &8.2$_{-1.2}^{+1.6}$ & 1.4$_{-0.2}^{+0.1}$ & 227/260\\ 
3 & 2.19$_{-0.04}^{+0.08}$ & 54$_{-12}^{+54}$ &5.3$_{-0.7}^{+3.1}$ &1.4$_{-0.2}^{+0.2}$ & 292/273\\
4-1 & 1.50$_{-0.02}^{+0.10}$ & 200$_{-20}^{+10}$ &6.4$_{-0.5}^{+0.6}$ & 1.8$_{-0.6}^{+0.2}$ & 283/269\\  
4-2 & 1.57$_{-0.04}^{+0.05}$ & 140$_{-7}^{+2}$ &6.0$_{-0.3}^{+0.3}$ & 1.7$_{-0.1}^{+0.1}$ & 292/264\\  
4-3 & 1.49$_{-0.04}^{+0.02}$ & 140$_{-10}^{+60}$ &7.0$_{-0.7}^{+1.1}$ & 2.1$_{-0.2}^{+0.1}$ & 283/264\\  
5 & 1.63$_{-0.02}^{+0.01}$ & 300$_{-40}^{+10}$ &7.0$_{-0.6}^{+1.2}$ & 2.0$_{-0.1}^{+0.2}$ & 371/278\\  
\hline
\end{tabular}
\vspace*{-0.5cm}
\end{table}
To constrain the inclination angle, a first step was to fit all the data
sets simultaneously keeping the inclination angle constant between each
observation. We obtain a best fit with an inclination angle of $\sim$ 50
deg. in agreement with the best fit angle obtained with the
{\sc{diskline}} model in the previous section (and also close to the
inclination angle found by Crummy et al. 2005 for OBS 3). The use of the
complete set of data was however too time consuming for the computation
of the errors. Thus, a second step was to study the different
observations separatly but fixing the inclination angle to the previous
value.

We obtain very good fits in most of the cases, with some problems with 
OBS 5 where important absorption features are present at low energies
(the fit is highly improved  
($\Delta\chi^2$ = 81) with the addition of a warm absorber component
{\sc absori} in {\sc xspec}). This model depends strongly on the 
emissivity law index $q$ ($j\propto r^{-q}$) and the inner disc radius
$r_{in}$. 
Thus the fit of the broad line
component as well as the soft excess requires a fine tuning of these two
parameters. For example, large $q$ and small $r_{in}$ imply strong
relativistic effects that strongly smeared the lines. This is
indeed what we find in the case of OBS 1, OBS 2 and OBS 5 where no broad
line is detected (cf. Table \ref{tabkdb}. On the other hand, for a
given $r_{in}$, smaller $q$ 
reduces the importance of the relativistic blurring effects and the broad
line is more easily detectable like in OBS 3 and OBS 4. The smaller
values of $q$ are obtained for OBS 3 and part 2 of OBS 4 where the
stronger broad lines are detected.

It is worth noting that in almost all cases but OBS 3 the addition of a
narrow line is required by the data. This is due to the large blurring
effects needed to fit the soft excess that smeared the iron line and
then prevent a good fit of the narrow component. With this model the soft band variability is due to both blurred reflection
and continuum variability while the hard band (above 3 keV) is dominated
by the power law continuum.
This model predicts however large reflection component  above 10 keV
i.e. outside the XMM 
energy range. In the case of the 2001 observations, we have 100\ ks
of simultaneous BeppoSAX observation, the BeppoSAX observation window
including OBS 1, 
OBS 2 and OBS 3. The BeppoSAX data show quite good agreements with the
XMM best fit models obtained in 2001 (cf. Fig. \ref{xmmsax}) which is
quite encouraging.  \vspace*{-0.25cm}

\section{Conclusion}
\vspace*{-0.25cm}
The present analysis of the complete set of XMM/EPIC-pn data of Mkn 841 reveals extreme and puzzling spectral and
temporal behaviors of its soft excess and iron line complex.
If the soft excess and broad iron line can be well fitted by an ionized
blurred reflection model, the origin of the narrow line component is
still not understood. Future Suzaku observations (2$\times$ 50 ks) will be of
crucial help to conclude on this source.


\vspace*{-0.25cm}

\begin{thebibliography}{}
\vspace*{-0.25cm}
  \bibitem{} Arnaud, K., Branduardi-Raymont, G., Culhane, J., et
  al.: 1985, MNRAS~217, 105 
  \bibitem{} Crummy J., Fabian, A., Brandt, W., Boller, Th.:
  2006, MNRAS~365, 1067
  \bibitem{} Falco, E., Kurtz, M., Geller, M., et al.: 1999, PASP~111, 438
  \bibitem{} George, I., Nandra, K., Fabian, A., et al. 1993: MNRAS~260, 111 
  \bibitem{} Iwasawa, K., Miniutti, G., Fabian, A.: 2004, MNRAS~355, 1073
  \bibitem{} Johnson, N.: 1997, BAAS~190, 2005
  \bibitem{} Longinotti, A., Nandra, K., Petrucci, P., O'Neill, P.:
  2004, MNRAS~355, 929 
  \bibitem{} Nandra, K., Turner, T., George, I., et al.: 1995, MNRAS~273, 85 
  \bibitem{} Petrucci, P., Henri, G., Maraschi, L., et al.: 2002,
  A\&A~388, L5  
  \bibitem{} Ross R., \& Fabian, A.: 2005, MNRAS~358, 211 

\end{thebibliography}
\end{document}